\documentclass[aps,showpacs,floatfix,twocolumn,pra]{revtex4}

\usepackage{epsfig,amsmath,graphicx,graphics,float}
\usepackage{cases,multimedia,bm,amssymb}

\begin{document}

\title{Anisotropic dynamics of a spin-orbit coupled Bose-Einstein
condensate}

\author{Giovanni I. Martone$^1$}
\author{Yun Li$^1$}
\author{Lev P. Pitaevskii$^{1,2}$}
\author{Sandro Stringari$^1$}

\affiliation{$^1$Dipartimento di Fisica, Universit\`{a} di Trento
and INO-CNR BEC Center, I-38123 Povo, Italy}
\affiliation{$^2$Kapitza Institute for Physical Problems RAS,
Kosygina 2, 119334 Moscow, Russia}

\begin{abstract}
By calculating the density response function we identify the
excitation spectrum of a Bose-Einstein condensate with equal Rashba
and Dresselhaus spin-orbit coupling. We find that the velocity of
sound along the direction of spin-orbit coupling is deeply quenched
and vanishes when one approaches the second-order phase transition
between the plane wave and the zero momentum quantum phases. We also
point out the emergence of a roton minimum in the excitation
spectrum for small values of the Raman coupling, providing the onset
of the transition to the stripe phase. Our findings point out the
occurrence of a strong anisotropy in the dynamic behavior of the
gas. A hydrodynamic description accounting for the collective
oscillations in both uniform and harmonically trapped gases is also
derived.
\end{abstract}


\maketitle

\section{Introduction}

Synthetic gauge fields are a developing field of research in atomic
physics. They have been the object of recent experimental
\cite{Lin2009, Lin2011, Aidelsburger2011, LeBlanc2012, Chen2012,
Zwierlein2012} and theoretical works \cite{Dalibard2010, Liu2009,
SO_2D_GS, SO_2D_GS_Trap, Ho2011, Li2012, Ozawa2012, Vyasanakere2011,
Gong2011}, giving rise to the occurrence of new quantum phases
exhibiting unique magnetic features, including spin-orbit coupled
configurations. The elementary excitations of such systems are also
expected to exhibit novel properties \cite{Bijl2011, Zhang2012,
Ramachandhran2012, ChenZhu2012, trento2, Zheng2012}. Some of these
features have been already the object of experimental measurements
\cite{Chen2012}. In particular the experiment of \cite{Chen2012} has
shown that the center-of-mass oscillation of a harmonically trapped
Bose-Einstein condensate can be deeply affected by the coupling with
the spin degree of freedom, in agreement with the predictions of
theory \cite{trento2}.

The purpose of the present work is to study the elementary
excitations and the corresponding behavior of the dynamic structure
factor of a spin-orbit coupled Bose-Einstein condensate (BEC) at
zero temperature, by direct investigation of the response of the gas
to a time-dependent perturbation. We explore both the phonon regime
of long wavelengths and the region at higher momentum transfer,
where the spectrum  exhibits novel features, including the
occurrence of a roton minimum. Our results point out the occurrence
of a strong anisotropy in the dynamic behavior of the gas.  In
ultracold gases the excitation spectrum can be measured via
two-photon Bragg spectroscopy \cite{steinhauer02}, so our
predictions can be relevant for future experiments on spin-orbit
coupled BECs.

\section{The Hamiltonian and the quantum phases}

We consider a spin $1/2$ Bose gas of $N$ particles enclosed in a
volume $V$, characterized by the single-particle Hamiltonian (we set
$\hbar=m=1$)
\begin{equation}
\begin{aligned}
h_0 = &\,\frac{{\bf p}^2}{2} +\frac{\Omega}{2}\sigma_x \cos(2k_0 x -
\Delta\omega_Lt)\\
&+\frac{\Omega}{2}\sigma_y\sin(2k_0x-\Delta\omega_Lt) -
\frac{\omega_Z}{2} \sigma_z
\end{aligned}
\label{eq:h00}
\end{equation}
accounting for the presence of two laser fields with frequencies
$\omega_L$ and $\omega_L+\Delta\omega_L$, wave vector difference
${\mathbf k}_0 = k_0 \hat{\mathbf e}_x$ along the $x$-direction, and
orthogonal linear polarizations providing transitions between the
two spin states via the Raman coupling $\Omega$. $\omega_Z$ is the
Zeeman shift between the two spin states in the absence of Raman
coupling \cite{Lin2011}, while $\sigma_k$, with $k=x,\,y,\,z$, are
the usual $2\times 2$ Pauli matrices. The Hamiltonian (\ref{eq:h00})
is not translationally invariant, but exhibits a screw-like
symmetry, being invariant with respect to helicoidal translations of
the form $e^{id(p_x-k_0\sigma_z)}$, consisting of a combination of a
rigid translation by distance $d$ and a spin rotation by angle $-d
k_0$ around the $z$-axis.

Let us now apply the unitary transformation $e^{i\Theta
\sigma_z/2}$, corresponding to a position and time-dependent
rotation in spin space by the angle $\Theta =2 k_0 x -
\Delta\omega_L t$, to the wave function obeying the Schr\"{o}dinger
equation. As a consequence of the transformation, the
single-particle Hamiltonian (\ref{eq:h00}) is transformed into the
translationally invariant and time-independent form
\begin{equation}
h_0^{\text{SO}}= \frac{1}{2}\left[\left(p_x-k_0 \sigma_z\right)^2+
p_\perp^2\right] + \frac{\Omega}{2} \sigma_x + \frac{\delta}{2}
\sigma_z \label{eq:h0}
\end{equation}
The spin-orbit nature acquired by the Hamiltonian results from the
non commutation of the kinetic energy and the position-dependent
rotation, while the renormalization of the effective magnetic field
$\delta = \Delta\omega_L-\omega_Z$ results from the additional time
dependence exhibited by the wave function in the rotating frame. The
new Hamiltonian is characterized by equal contributions  of Rashba
\cite{Rashba1984} and Dresselhaus \cite{Dresselhaus1955} couplings.
It is worth pointing out that the operator $\mathbf{p}$ entering
(\ref{eq:h0}) is the canonical momentum $-i\nabla$, the physical
velocity being given by $\mathbf{v}_\pm=\mathbf{p} \mp
k_0\hat{\mathbf e}_x$ for the spin-up and spin-down particles. In
terms of $\mathbf{p}$ the eigenvalues of (\ref{eq:h0}) are given by
(we set here $\delta=0$)
\begin{equation}
\epsilon_{\pm}(\mathbf{p}) = \frac{p_x^2 + p_\perp^2 + k_0^2}{2} \pm
\sqrt{k_0^2 p_x^2 + \frac{\Omega^2}{4}} \label{eq:epsilonp}
\end{equation}
and are characterized by a double band structure.

In the presence of two-body interactions the Hamiltonian of the
$N$-body system is given by
\begin{equation}
H = \sum_j h_0^{\text{SO}}(j) + \sum_{\alpha,\,\beta} \frac{1}{2}
\int d^3\mathbf{r} \, g_{\alpha\beta}\, n_\alpha(\mathbf{r})
n_\beta(\mathbf{r}), \label{eq:Ham}
\end{equation}
where $h_0^{\text{SO}}$ is given by (\ref{eq:h0}), and
$\alpha,\beta$ are the spin indices ($\uparrow,\downarrow \,=\pm$)
characterizing the two spin states. The spin-up and spin-down
density operators entering Eq.~(\ref{eq:Ham}) are defined by
$n_\pm({\mathbf r})= (1/2) \sum_j \left(1 \pm \sigma_{z,\,j} \right)
\delta({\mathbf r}-{\mathbf r}_j)$, while $g_{\alpha\beta} = 4\pi
a_{\alpha\beta}$ are the relevant coupling constants in the
different spin channels, with $a_{\alpha\beta}$ the corresponding
$s$-wave scattering lengths. Notice that the two-body interaction
terms are not affected by the spin rotation discussed before.

\begin{figure}
\centering
\includegraphics[scale=0.36]{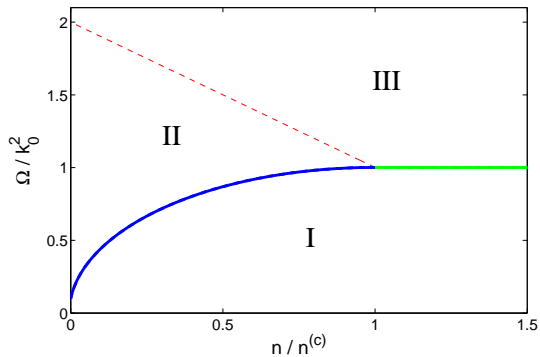}
\caption{(Color online) Phase diagram corresponding to the
spin-orbit coupled Hamiltonian (\ref{eq:Ham}). The lines
corresponding to the I-II (blue), II-III (red) and I-III (green)
phase transitions are shown. The parameters: $g = 4\pi \times 100 \,
a_B$, where $a_B$ is the Bohr radius, $\gamma=0.0012$, $k_0^2 = 2\pi
\times 80$ Hz, corresponding to a critical density $n^{(c)} =
k_0^2/(2\gamma g) = 4.37 \times 10^{15}$ cm$^{-3}$.}
\label{fig:phase_diag}
\end{figure}

The Hamiltonian (\ref{eq:Ham}) has been already implemented
experimentally \cite{Lin2011, Chen2012} and has been recently
employed to predict a variety of nontrivial quantum phases in
Bose-Einstein condensates \cite{Ho2011, Li2012}. It has the peculiar
property of violating both parity and time-reversal symmetry. In the
presence of a spin symmetric interaction ($g_{\uparrow \uparrow} =
g_{\downarrow\downarrow} = g$ and $\delta=0$), the quantum phases
predicted by mean-field theory depend on the value of the relevant
parameters $k_0$, $\Omega$, and the interaction parameters
\cite{note_g}
\begin{equation}
G_1 = n\left(g + g_{\uparrow\downarrow}\right)/4\,, \qquad G_2 =
n\left(g - g_{\uparrow\downarrow}\right)/4  \label{eq:G12}
\end{equation}
where $n = N/V$ is the average density. In uniform matter one can
use the ansatz
\begin{equation}
\psi= \sqrt{n} \left[C_+\begin{pmatrix} \cos \theta \\
-\sin\theta\end{pmatrix} e^{ik_1x}+C_-\begin{pmatrix} \sin \theta \\
-\cos\theta\end{pmatrix} e^{-ik_1x} \right] \label{eq:spinor0}
\end{equation}
for the ground state wave function of the condensate, with
$|C_+|^2+|C_-|^2=1$, and $k_1$ representing the momentum where
Bose-Einstein condensation takes place. Energy minimization with
respect to $k_1$ yields the general relationship $\theta=
\arccos(k_1 /k_0)/2$ fixed by the single-particle Hamiltonian
(\ref{eq:h0}). Minimization with respect to the other parameters
eventually permits to calculate key physical quantities like the
momentum distribution and the longitudinal
($\langle\sigma_z\rangle$) and transverse ($\langle\sigma_x
\rangle$, $\langle\sigma_y\rangle$) spin polarization of the gas
\cite{sigmaz}:
\begin{eqnarray}
\langle\sigma_z\rangle &\hspace{-1mm}=&\hspace{-1mm}
\left(|C_+|^2-|C_-|^2\right) \frac{k_1}{k_0} \label{eq:sigma_z}\\
\langle\sigma_x\rangle &\hspace{-1mm}= &\hspace{-1mm} -\left[
\frac{\sqrt{k_0^2 -k_1^2}}{k_0} +2|C_+ C_-|\cos \left(2k_1 x+
\phi\right)\right] \;\; \label{eq:sigma_x}\\
\langle\sigma_y\rangle &\hspace{-1mm}= &\hspace{-1mm} |C_+ C_-|
\frac{2 k_1}{k_0}\sin\left(2k_1 x +\phi\right) \label{eq:sigma_y}
\end{eqnarray}
where $\langle\;\rangle$ corresponds to the average in spin space
divided by the average density $n$, and $\phi$ is the relative phase
between $C_+$ and $C_-$. The resulting ground state for $G_1>0$ is
compatible with the three distinct BEC phases (see Fig.
\ref{fig:phase_diag}).

{\bf Phase I}.  For small values of the Raman coupling $\Omega$, and
positive values of $G_2$, the ground state corresponds to a linear
combination of the two plane waves $e^{\pm i k_1 x}$  with equal
weight ($|C_+|=|C_-|=1/\sqrt{2}$). This phase (hereafter called
stripe phase or phase I) shares important analogies with
supersolids, being characterized by the co-existence of BEC and  by
density modulations in the form of stripes, whose actual spatial
location is the result of a mechanism of spontaneous  breaking of
translational invariance. The density modulations take the form
$n(\mathbf{r}) = n [1 + \sqrt{1-(k_1/k_0)^2} \cos(2 k_1 x + \phi)]$,
with $k_1= k_0\sqrt{1 -\Omega^2/[2(k^2_0+G_1)]^2}$. It is worth
mentioning that these modulations differ from the ones of the laser
potential (see Eq.~(\ref{eq:h00})) and have a different nature with
respect to the modulations exhibited by the density in the presence
of usual optical lattices. The contrast in $n({\bf r})$ vanishes as
$\Omega \to 0$ as a consequence of the orthogonality of the two spin
states (in fact in this limit  $\theta = 0$ and  $k_1 = k_0$). In
the stripe phase the longitudinal spin density identically vanishes:
$\langle\sigma_z\rangle=0$, while $\langle\sigma_x\rangle \neq 0$.
It is worth mentioning that the ansatz (\ref{eq:spinor0}) for the
stripe phase provides only a first approximation which ignores
higher-order harmonics caused by the nonlinear interaction terms in
the Hamiltonian.

{\bf Phase II}. For larger values of the Raman coupling the system
enters a new phase, the so-called plane wave phase (hereafter called
phase II), where BEC takes place in a single plane wave state with
momentum  $\mathbf{p}=k_1\hat{\mathbf e}_x$, lying on the $x$-axis
(in the following we choose $k_1>0$). In this phase, the density is
uniform. The spin polarization characterizing this phase is given by
the simple expression $\langle\sigma_z\rangle = k_1/k_0$, with $k_1=
k_0\sqrt{1- \Omega^2/[2\left(k^2_0-2G_2\right)]^2}$, while the
transverse polarization is given by $\langle \sigma_x \rangle = -
\Omega/[2(k^2_0-2G_2)]$. An energetically equivalent configuration
is obtained by considering the BEC in the single-particle state with
$\mathbf{p}=-k_1 \hat{\mathbf e}_x$, the choice between the two
configurations being determined by a mechanism of spontaneous
symmetry breaking, typical of a ferromagnetic configuration.

{\bf Phase III}. At even larger values of $\Omega$ the system enters
the so-called zero momentum phase (phase III), where the condensate
has zero momentum ($k_1=0$), the density is uniform, and the
longitudinal spin polarization $\langle \sigma_z \rangle$
identically vanishes, while $\langle\sigma_x \rangle =-1$.

The chemical potential in the three phases can be calculated
following the procedure of \cite{Li2012} and is given by
\begin{eqnarray}
&& \mu^{(\text{I})} = 2G_1 -\frac{k_0^2 \Omega^2}{8 \left(k_0^2+
G_1 \right)^2} \label{eq:mu_I}\\
&& \mu^{(\text{II})} = 2 \left(G_1+G_2\right) -\frac{k_0^2
\Omega^2}{8 \left(k_0^2 -2G_2\right)^2} \label{eq:mu_II} \\
&& \mu^{(\text{III})} = 2G_1 + \frac{k_0^2 - \Omega}{2}.
\label{eq:mu_III}
\end{eqnarray}
The critical values of the Raman frequencies $\Omega$ characterizing
the phase transitions are obtained by imposing that the chemical
potential and the pressure $P=n\mu(n) -  \int\mu(n)dn$ be equal in
the two phases at equilibrium. One finds that the transition between
phases I and II is of first-order nature and is characterized by
different values of the densities of the two phases. The density
differences are however extremely small and are not visible in Fig.
\ref{fig:phase_diag}. The transition between phases II and III is
instead of second-order nature and is characterized by a jump in the
compressibility $n^{-1}(\partial \mu / \partial n)^{-1}$ if $G_2\ne
0$, and by a divergent behavior of the spin polarizability (see
Section IV). For small values of the coupling constants ($G_1, G_2
\ll k^2_0$) the critical value of the Raman coupling
$\Omega^{(\text{I-II})}$ between phases I and II is given by the
density-independent expression \cite{Ho2011,Li2012}
\begin{equation}
\Omega^{(\text{I-II})} =2 k_0^2 \sqrt{\frac{2\gamma}{1+2\gamma}}
\label{eq:OmegaI-II}
\end{equation}
with $\gamma = G_2/G_1$. The transition between phases II and III
instead takes place at the higher value \cite{Li2012}
\begin{equation}
\Omega^{(\text{II-III})}=2\left(k_0^2 - 2 G_2\right)
\label{eq:OmegaII-III}
\end{equation}
provided that the condition $k^2_0 > 4G_2\left(1+\gamma\right)$ is
satisfied; in the opposite case one instead has the first-order
transition directly between phases I and III \cite{Li2012}.  One
should finally remind that if $G_2 < 0$ only phases II and III are
available, the stripe phase being always energetically unfavorable.

\section{Density response function}

In order to calculate the dynamic density response of the system we
add the time-dependent perturbation $V_{\lambda}= - \lambda
e^{i(\mathbf{q}\cdot \mathbf{r}-\omega t)}+\text{H.c.}$ to the
single-particle Hamiltonian (\ref{eq:h0}). The direction of the wave
vector $\mathbf{q}$ is characterized by the polar angle $0\le \alpha
\le \pi$ with respect to the $x$-axis. The density response function
is then calculated through the usual definition $\chi(\mathbf{q},
\omega) = \lim_{\lambda \to 0}\delta \rho_{\bf q} /(\lambda
e^{-i\omega t})$, where $\delta \rho_{\bf q}$ are the fluctuations
of the ${\bf q}$-component of the density induced by the external
perturbation. In the following we calculate $\chi(\mathbf{q},
\omega)$ by solving the time-dependent Gross-Pitaevskii equation
\begin{equation}
i \partial_t \psi = \left[h_0^{\text{SO}} + V_\lambda + \frac{2
G_1}{n} \left(\psi^\dagger \psi\right) + \frac{2 G_2}{n}
\left(\psi^\dagger \sigma_z \psi\right)\sigma_z\right]\psi
\label{eq:spinorGP}
\end{equation}
where $h_0^{\text{SO}}$ is the single-particle Hamiltonian
(\ref{eq:h0}) with $\delta=0$. We restrict the analysis to phases II
and III, where the ground state density is uniform and the wave
function of the condensate can be written in the simple form
\begin{equation}
\begin{split}
\psi= \sqrt{n} & \left[\begin{pmatrix} \cos \theta \\
-\sin\theta\end{pmatrix} e^{ik_1x} \right.\\
& \left.+ \begin{pmatrix}u_\uparrow(\mathbf{r}) \\
u_\downarrow(\mathbf{r}) \end{pmatrix} e^{-i \omega t} +
\begin{pmatrix} v_\uparrow^\ast(\mathbf{r})\\
v_\downarrow^\ast(\mathbf{r}) \end{pmatrix} e^{i \omega t}\right]
e^{- i \mu t}.
\end{split} \label{eq:spinor}
\end{equation}
The terms depending on the Bogoliubov amplitudes $u$ and $v$ provide
the deviations in the order parameter with respect to equilibrium,
caused by the external perturbation. In the linear, small $\lambda$,
limit we find the result (near the poles one should replace $\omega$
with $\omega +i0$)
\begin{widetext}
\begin{equation}
\chi(\mathbf{q},\omega)= \frac{- N q^2\left[ \omega^2 - 4 k_1 q
\cos\alpha \, \omega + a(q,\,\alpha)\right]} {\omega^4 -4 k_1 q
\cos\alpha \, \omega^3 + b_2(q,\,\alpha) \omega^2 + k_1 q \cos\alpha
\, b_1(q,\,\alpha) \omega + b_0(q,\,\alpha)} \label{eq:response}
\end{equation}
\end{widetext}
where the coefficients $a$ and $b_i$ are even functions of $q \equiv
|\mathbf{q}|$ and $\cos\alpha$, implying that $b_i(q,\,\alpha)
=b_i(q,\, \pi \pm\alpha)$ (the same for $a$), and their actual
values depend on whether one is in phase II or III (see Appendix A).

The above equations include all the relevant information relative to
the frequency of the elementary excitations, given by the poles of
$\chi$, i.e. by the zeros of
\begin{equation}
\omega^4 - 4 k_1 q \cos\alpha \, \omega^3 + b_2 \omega^2 + k_1 q
\cos\alpha \, b_1 \omega + b_0 = 0, \label{eq:dispersion}
\end{equation}
as well as to the dynamic structure factor given, at $T=0$, by
\begin{equation}
S(\mathbf{q},\,\omega)= \pi^{-1}\text{Im} \chi(\mathbf{q}, \,\omega)
\label{eq:Sq}
\end{equation}
for $\omega \ge 0$ and $S(\mathbf{q},\,\omega)=0$ for negative
$\omega$. In particular the $f$-sum rule $\int d\omega
S(\mathbf{q},\,\omega)\omega=N q^2/2$ is exactly satisfied, as one
can deduce from the correct large $\omega$  behavior of the density
response function: $\chi(\mathbf{q},\,\omega)_{\omega\to\infty} = -
N q^2/\omega^2$ \cite{book}. It is also worth pointing out that the
density response function is invariant with respect to the unitary
transformation yielding the Hamiltonian in the spin-rotated frame,
so that the results presented in this paper, based on
Eq.~(\ref{eq:response}), hold also in the original frame and are
relevant for actual experiments.

Equation (\ref{eq:response}) reduces to a simplified form in two
limiting cases. A first case is when $G_2=0$ and $\Omega=0$. In this
limit the denominator can be rewritten in a factorized form and
$\chi$ reduces to the usual Bogoliubov form $\chi(\mathbf{q},\,
\omega)= - N q^2/[\omega^2 - q^2( 2G_1 +q^2/4)]$ characterizing the
response of a BEC gas in the absence of spin-orbit coupling. A
second case is the ideal Bose gas ($G_1=G_2=0$) where $H$ reduces to
the single-particle Hamiltonian (\ref{eq:h0}) with $\delta=0$ and
the excitation spectrum, given by the solutions of
Eq.~(\ref{eq:dispersion}), takes the simple form:
\begin{equation}
\omega_\pm(\mathbf{q}) = \epsilon_\pm(\mathbf{p}_1 + \mathbf{q}) -
\epsilon_-(\mathbf{p}_1) \label{eq:omegapm}
\end{equation}
where $\mathbf{p}_1=k_1\hat{\mathbf e}_x$ is the momentum where
Bose-Einstein condensation takes place, and $\epsilon_\pm$ are the
two branches of the single-particle spectrum (\ref{eq:epsilonp}).

It is worth noticing that the odd terms in $\omega$ entering the
response function identically vanish in the zero momentum phase III,
but survive in phase II, reflecting the lack of parity and time
reversal symmetry of the ground state wave function. The condition
$\text{Im}\chi(\mathbf{q},\,\omega)= -\text{Im} \chi(-\mathbf{q},\,
-\omega)$, characterizing the imaginary part of the response
function, is always satisfied, but the symmetry relationship
$\text{Im} \chi(\mathbf{q},\,\omega) =\text{Im} \chi(-\mathbf{q},
\omega)$ is not ensured in phase II, where one consequently finds
$S(\mathbf{q},\,\omega)\ne S(-\mathbf{q},\,\omega)$. First results
for the excitation spectrum of the Hamiltonian (\ref{eq:Ham}) for
small and large values of $\Omega$, far from the transition between
the plane wave and the zero momentum phases, have been recently
discussed in \cite{Zheng2012} using a hydrodynamic formalism.

Equation (\ref{eq:response}) permits to calculate the static
response function $\chi(\mathbf{q}) \equiv \chi(\mathbf{q},\,
\omega=0)/N$ yielding the results
\begin{eqnarray}
\mathcal{K}^{-1}_{\text{II}} &\hspace{-1mm}=&\hspace{-1mm} 2 G_1 +
\frac{2 G_2 k_1^2\left(k_1^2\cos^2\alpha + k_0^2\sin^2\alpha - 2
G_2\right)}{k_1^2\left(k_0^2\cos^2\alpha - 2 G_2\right) +
k_0^4\sin^2\alpha} \qquad \label{eq:compress_II}\\
\mathcal{K}_{\text{III}}^{-1} &\hspace{-1mm}=&\hspace{-1mm} 2 G_1
\label{eq:compress_III}
\end{eqnarray}
for the $q=0$ value $\mathcal{K}\equiv \chi(q=0)$ of the static
response in phases II and III respectively. The result
(\ref{eq:compress_II}) depends on the polar angle $\alpha$,
revealing the anisotropy of $\mathcal{K}$ in the plane wave phase
caused by the spin interaction term $G_2$. It is also worth pointing
out that, if $\cos\alpha\neq \pm 1$, in phase II the $q=0$ static
response $\mathcal{K}$ differs from the thermodynamic
compressibility $n^{-1}(\partial \mu / \partial n)^{-1}$ with $\mu$
calculated from (\ref{eq:mu_II}). Furthermore, if $\cos\alpha =\pm
1$ and $G_2 \ne 0$, the $q=0$ static response $\mathcal{K}$ exhibits
a jump at the transition between phases II and III. One can easily
prove that the frequencies $\omega(\mathbf{q})$ of the elementary
excitations, given by the zeros of (\ref{eq:dispersion}), are
instead always continuous functions of the Raman coupling $\Omega$
at the transition for all values of $\mathbf{q}$.

\section{Velocity of sound and the role of the magnetic susceptibility}

\begin{figure}[b]
\includegraphics[scale=0.97]{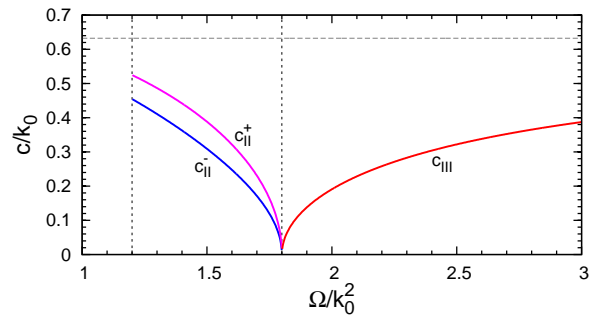}
\caption{(Color online) Sound velocity as a function of the Raman
coupling for the following choice of parameters: $G_1/k_0^2 = 0.2$,
$G_2/k_0^2 = 0.05$. The two sound velocities in phase II correspond
to phonons propagating in the direction parallel ($c^+_{\text{II}}$)
and antiparallel ($c^-_{\text{II}}$) to $k_1$. The horizontal dashed
line corresponds to the value $\sqrt{2G_1}=0.63\,k_0$ of the sound
velocity in the absence of spin-orbit and Raman coupling. The
vertical dashed lines indicate the Raman frequencies at which the
I-II and II-III phase transitions take place.}
\label{fig:sound_velocity}
\end{figure}

The low frequency excitations at small $q$ (sound waves) can be
easily obtained by setting $\omega=c q$  and keeping the leading
terms in $q^2$ in (\ref{eq:dispersion}). In phase III we find the
result
\begin{equation}
c_{\text{III}}=\sqrt{2 G_1\left(1 - \frac{2 k_0^2 \cos^2
\alpha}{\Omega + 4 G_2}\right)} \label{eq:c_III}
\end{equation}
which explicitly shows the strong reduction of the sound velocity
along the $x$-direction ($\cos\alpha=\pm 1$) caused by the
spin-orbit coupling when one approaches the transition to the plane
wave phase. The quenching can be understood in terms of the increase
of the effective mass associated with the single-particle spectrum
(\ref{eq:epsilonp}). At the transition, where the velocity of sound
propagating along the $x$-direction vanishes, the elementary
excitations exhibit a different $q^2$ dependence. On the other hand,
the sound velocities along the other directions ($\alpha \neq 0$ and
$\pi$) remain finite at the transition. In the plane wave phase II,
the sound velocity is instead given by
\begin{widetext}
\begin{equation}
c_{\text{II}} = \frac{\sqrt{
2\left[G_1 k_0^4 + G_2 k_1^2 \left(k_0^2 - 2 G_1 - 2 G_2\right)\right]
\left[k_0^4 - 2 G_2 k_1^2 - k_0^2 \left(k_0^2 - k_1^2\right) \cos^2\alpha
\right]} + 2G_2 k_1\left(k_0^2 - k_1^2\right) \cos \alpha}{k_0^4 - 2
G_2 k_1^2} \label{eq:c_II}
\end{equation}
\end{widetext}
and exhibits a further interesting feature caused by the lack of
parity symmetry. The asymmetry effect in the sound velocity is due
to the presence of the last term in the numerator of
Eq.~(\ref{eq:c_II}), therefore the symmetry will be recovered if
$G_2=0$ or $\alpha=\pi/2$ (corresponding to phonons propagating
along the directions orthogonal to the $x$-axis). Also in phase II,
the velocity of sound along the $x$-direction vanishes when one
approaches the transition to phase III.

In order to better understand the role played by the spin degree of
freedom in the propagation of sound, it is interesting to relate the
sound velocity to the magnetic polarizability, which can be
calculated by generalizing the ground state condensate wave function
(\ref{eq:spinor0}) in the presence of a static magnetic field $h$
coupled to the system through the interaction term $-h\sigma_z$. To
calculate the new ground state we replace the variational parameters
$\theta$ and $k_1$ entering the ansatz (\ref{eq:spinor0}) with two
independent sets of parameters $\theta_+$, $k_1^+$ and $\theta_-$,
$k_1^-$ characterizing the two plane waves, and minimize the energy.
In the small $h$ limit the magnetic polarizability is determined by
$\mathcal{M}=\int d^3 r\langle\sigma_z\rangle/(h V)$. After some
straightforward algebra we find the following results holding,
respectively, in phases II and III \cite{trento2}:
\begin{eqnarray}
&&\mathcal{M}_{\text{II}} = \frac{k_0^2-k_1^2}{k_1^2 \left(k_0^2 -2
G_2\right)}\\
&&\mathcal{M}_{\text{III}} = \frac{2}{\Omega-2\left(k_0^2 - 2
G_2\right)}.
\end{eqnarray}
A peculiar feature exhibited by the above equations is their
divergent behavior near the second-order phase transition II-III
where $\Omega= 2(k^2_0-2G_2)$ and $k_1=0$. In terms of the $q=0$
static response $\mathcal{K}$ and the magnetic susceptibility
$\mathcal{M}$ one can rewrite the results for the sound velocity in
the useful form
\begin{equation}
c(\alpha) c(\alpha+\pi) = \frac{1+k_0^2 \mathcal{M} \sin^2
\alpha}{\mathcal{K}\left(1+k_0^2 \mathcal{M}\right)}
\label{eq:c+_c-}
\end{equation}
holding in both phases II and III. Equation (\ref{eq:c+_c-})
generalizes the usual relation $c^2=n(\partial\mu/ \partial n)$
between the sound velocity and the compressibility holding in usual
superfluids. It explicitly shows that, along the $x$-direction,
where $\sin\alpha=0$, the sound velocity $c$ vanishes at the
transition because of the  divergent behavior of the magnetic
polarizability. The results for the sound velocity along the
$x$-axis are shown in Fig.~\ref{fig:sound_velocity} for a
configuration with relatively large $G_2$, emphasizing the
difference between $c^+_{\text{II}}(\alpha=0)$ and $c^-_{\text{II}}
(\alpha = \pi)$, i.e between the velocities of sound waves
propagating in opposite directions along the $x$-axis. Notice that
the sound velocity, in the absence of spin-orbit and Raman coupling,
would correspond to the value $c=\sqrt{2 G_1}$ (horizontal dashed
line). This value is asymptotically reached only for very large
values of $\Omega$. The quenching effect exhibited by the sound
velocity near the II-III phase transition is particularly remarkable
in the zero momentum phase III where BEC takes place in the
$\mathbf{p}=0$ state and the compressibility of the gas is
unaffected by spin-orbit coupling. It explicitly reveals the mixed
density and magnetic nature of the sound waves, the spin nature
becoming more and more important as one approaches the phase
transition where $\mathcal{M}$ diverges.

It is finally interesting to understand the role played by the sound
waves in terms of sum rules. From Eq.~(\ref{eq:response}) one can
easily prove that phonons exhaust the compressibility sum rule
$\int_{-\infty}^{+\infty} d\omega S(\mathbf{q}, \omega)/\omega$ at
small $q$ but, differently from ordinary superfluids, they give only
a small contribution to the $f$-sum rule $\int_{-\infty}^{+\infty}
d\omega S(\mathbf{q}, \omega)\omega= N q^2/2$ as one approaches the
transition \cite{note1}. This contribution becomes vanishingly small
at the transition for wave vectors $\mathbf{q}$ oriented along the
$x$-direction. Also the static structure factor $S(\mathbf{q})=
\int_0^\infty d\omega S(\mathbf{q}, \omega)/N$ is strongly quenched
compared to usual BECs. This results in an enhancement of the
quantum fluctuations of the order parameter as predicted by the
uncertainty principle inequality \cite{uncertainty}. The effect is
however small because the sound velocity vanishes only along the
$x$-direction \cite{Li2012}.

\section{Roton and Maxon excitations}

\begin{figure}
\centering
\includegraphics[scale=0.9]{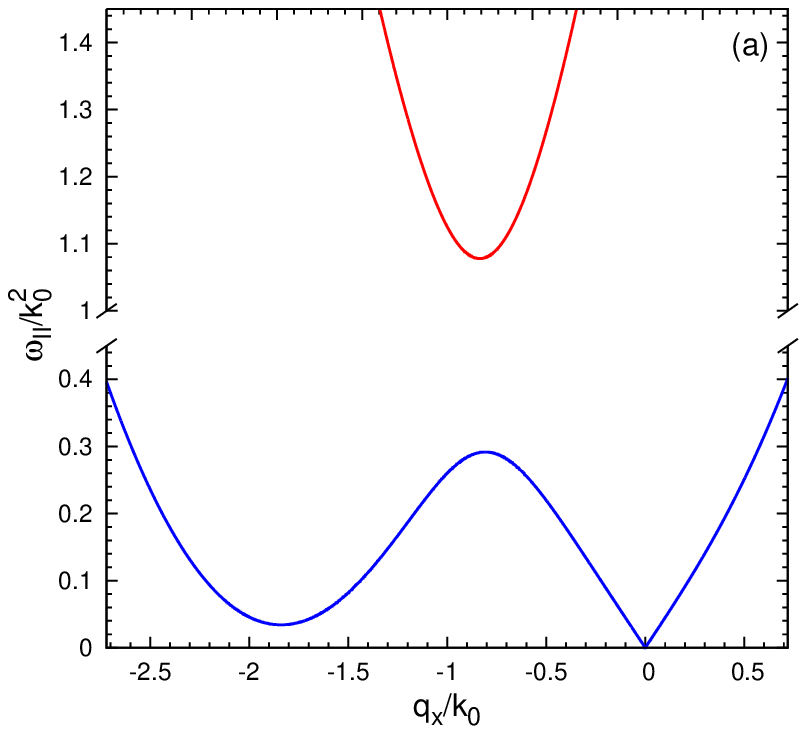}
\includegraphics[scale=0.9]{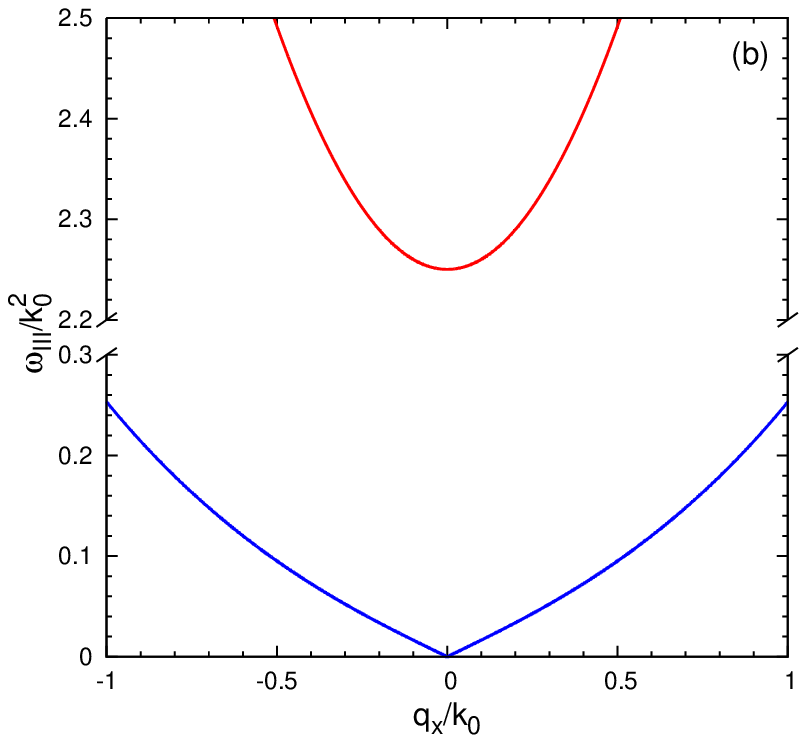}
\caption{(Color online) Excitation spectrum in phase II (a)
($\Omega/k_0^2=0.85$) and in phase III (b) ($\Omega/k_0^2=2.25$) as
a function of  $q_x$ ($q_y=q_z=0$). The blue and red lines represent
the lower and upper branches, respectively. In phase II the spectrum
is not symmetric and exhibits a roton minimum for negative $q_x$,
whose energy becomes smaller and smaller as one approaches the
transition to the stripe phase at $\Omega/k_0^2=0.09$. The other
parameters: $G_1/k_0^2=0.12$, $\gamma = G_2/G_1= 10^{-3}$.}
\label{fig:spectrum}
\end{figure}

When one moves far from the phonon regime new interesting features
emerge from the study of the response function. First the poles of
Eq.~(\ref{eq:response}) provide two separated branches (see
Figs.~\ref{fig:spectrum}a and \ref{fig:spectrum}b), the lower one
approaching the phonon dispersion at small $q$. For example in phase
III, where the excitation spectrum is symmetric under inversion of
$\mathbf{q}$ into $-\mathbf{q}$, the gap between the two branches is
given, at $\mathbf{q}=0$, by $\Delta = \sqrt{\Omega(\Omega + 4
G_2)}$.

\begin{figure}
\centering
\includegraphics[scale=0.9]{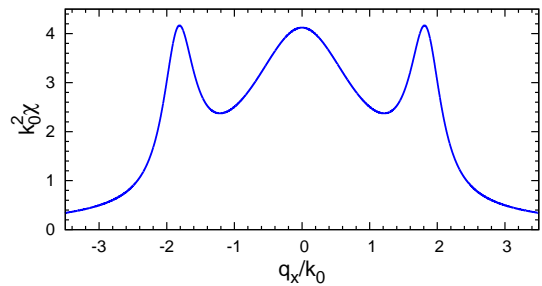}
\caption{(Color online) Static response in phase II as a function of
$q_x$ ($q_y=q_z=0$). The curve is symmetric and exhibits a typical
peak near the roton momentum. The parameters: $\Omega/k_0^2=0.85$,
$G_1/k_0^2=0.12$, $\gamma = G_2/G_1= 10^{-3}$.}
\label{fig:pw_statrespfunc}
\end{figure}

A very peculiar feature of the lower branch is exhibited in the
plane wave phase II for negative values of $q_x$, resulting in the
emergence of a roton minimum \cite{Zheng2012} which becomes more and
more pronounced as one approaches the phase transition to the stripe
phase I. The occurrence of the rotonic structure in spin-orbit
coupled BEC gases shares interesting analogies with the case of
dipolar gases in quasi 2D configurations \cite{Bohn2012}. In
Fig.~\ref{fig:spectrum}a we show the excitation spectrum in phase
II, calculated in the experimental conditions of \cite{Chen2012},
for wave vectors $\mathbf{q}$ lying on the $x$-axis. In
Fig.~\ref{fig:spectrum}b we instead show the excitation spectrum in
phase III which, differently from Fig.~\ref{fig:spectrum}a, exhibits
symmetry under inversion of $q_x$ into $-q_x$. The physical origin
of the roton minimum is quite clear. In phase II the ground state is
degenerate and it is very favorable for atoms to be transferred from
the BEC state at $\mathbf{p} = \mathbf{p}_1$ to the empty state at
$\mathbf{p} = -\mathbf{p}_1$. The occurrence of the roton minimum is
also reflected in a strong enhancement in the static response
function $\chi(q_x)$ (see Fig.~\ref{fig:pw_statrespfunc}). Notice
that $\chi(q_x)$, differently from $\omega(q_x)$, is always a
symmetric function of $q_x$. The occurrence of the roton minimum in
the excitation spectrum and the corresponding enhancement of the
static response represent a typical tendency of the system towards
crystallization. In the case of excitations propagating along the
$x$-axis we have investigated in detail the condition for the roton
frequency being equal to zero, corresponding to a divergent behavior
for $\chi(q_x)$. A simple analytic expression for the corresponding
value of the Raman coupling $\Omega$ is obtained in the weak
coupling limit $G_1, G_2 \ll k^2_0$ where we find that the critical
value exactly coincides with the value (\ref{eq:OmegaI-II})
characterizing the transition between the plane wave and the stripe
phases. For larger values of the coupling constants $G_1$ and $G_2$
we expect that the critical value takes place for values of the
Raman coupling smaller than the value at the transition, exhibiting
the typical spinoidal behavior of first-order liquid-crystal phase
transitions.

\begin{figure}
\centering
\includegraphics[scale=0.9]{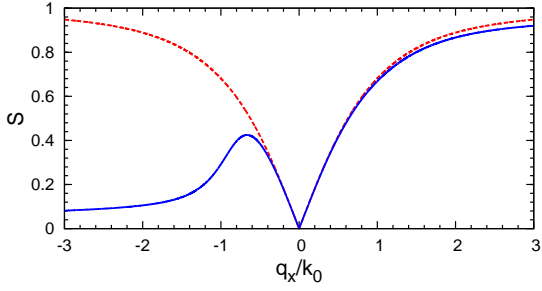}
\caption{(Color online) Contribution of the lower branch to the
static structure factor in phase II, as a function of $q_x$ (blue
solid line), compared with the total $S(q_x)$ (red dashed line). The
parameters: $\Omega/k_0^2=0.85$, $G_1/k_0^2=0.12$, $\gamma =
G_2/G_1= 10^{-3}$.} \label{fig:pw_statstrucfact}
\end{figure}

Despite the divergent behavior exhibited by the static response
function $\chi(q_x)$, the static structure factor $S(q_x)$ does not
exhibit any  peaked structure near the roton point, differently from
what happens, for example, in superfluid Helium \cite{T}. In
Fig.~\ref{fig:pw_statstrucfact} we show $S(q_x)$ together with the
contribution to the integral $S(q_x) = \int d\omega S(q_x,\omega)/N$
arising from the lower branch of the elementary excitations. In the
figure we have chosen $q_y=q_z=0$. The figure shows that the lower
branch contribution is not symmetric for exchange of $q_x$ into
$-q_x$, even if the total $S(q_x)$ is symmetric \cite{S+S-}.
Remarkably, the figure shows that the strength carried by the lower
branch is significantly  peaked for intermediate values of $q_x$
between the phonon and the roton regimes, in the so called maxon
region, where the lower branch excitation spectrum exhibits a
maximum (see Fig.~\ref{fig:spectrum}a).

\section{Hydrodynamic formalism}

The peculiar behavior of the excitation spectrum in the phonon
regime discussed in Sect. IV can be usefully described using the
hydrodynamic formalism where one writes the spin-up and spin-down
components of the order parameter in terms of their modulus and
phase \cite{Zheng2012}. In this case one finds four coupled
equations instead of two equations as in usual BECs. In the phonon
regime of large wavelengths and small frequencies one can safely
neglect the quantum pressure terms. Furthermore, one finds that the
phase difference between the two spin components is blocked
($\varphi_\uparrow = \varphi_\downarrow$). This is the consequence
of the equation for the spin density and the fact that $\omega \ll
\Omega$ \cite{note3}. By imposing the condition $\varphi\equiv
\varphi_\uparrow=\varphi_\downarrow$, holding for small frequencies,
one then derives the non trivial relationship
\begin{equation}
k_0\nabla_x \delta\varphi - k_0^2 Z \left(-\frac{s}{n}\frac{\delta
n}{n} + \frac{\delta s}{n}\right) - 2 G_2 \frac{\delta s}{n} = 0
\label{eq:newrelation}
\end{equation}
between the phase gradient, the density and the spin fluctuations.
In the above equation $s = n k_1/k_0$ is the spin density relative
to the equilibrium configuration and we have defined the relevant
parameter
\begin{equation}
Z = \frac{\Omega}{2 k_0^2 \left(1 - k_1^2/k_0^2\right)^{3/2}}.
\label{eq:Z}
\end{equation}
Equation (\ref{eq:newrelation}) permits to reduce the hydrodynamic
equations
\begin{eqnarray}
\partial_t \delta n &\hspace{-2mm}+&\hspace{-2mm} \nabla \cdot \left(n
\nabla\delta\varphi\right) - k_0 \nabla_x\left[ n\left( -\frac{s}{n}
\frac{\delta n}{n} + \frac{\delta s}{n} \right) \right] = 0,\;\;\;
\label{eq:continuity}\\
\partial_t \delta\varphi &\hspace{-2mm}+&\hspace{-2mm} k_1 \nabla_x
\delta\varphi  - k_0^2 Z\frac{s}{n} \left(-\frac{s}{n}\frac{\delta
n}{n} + \frac{\delta s}{n}\right) + 2 G_1 \frac{\delta n}{n} = 0\;\;
\nonumber\\
&& \label{eq:phase}
\end{eqnarray}
for the density and the phase, respectively, to a closed set of
coupled equations. The solutions of the hydrodynamic equations
reproduce exactly the results (\ref{eq:c_III}) and (\ref{eq:c+_c-})
for the sound velocity. It is in particular worth pointing out the
crucial changes caused by the spin-orbit term in the equation of
continuity (\ref{eq:continuity}). These changes reflect the fact
that the current is not simply given by the canonical momentum
operator, but is affected by the spin variable. The current density
operator should actually satisfy the continuity equation
$[H,n({\mathbf r})] = i \nabla \cdot {\mathbf j}({\mathbf r})$,
where $n({\mathbf r})= \sum_k \delta({\mathbf r}-{\mathbf r}_k)$ is
the density operator. By explicitly carrying out the commutator one
identifies the current as ${\mathbf j}({\mathbf r})= {\mathbf
p}({\mathbf r}) - k_0\sigma_z({\mathbf r}) \hat{\mathbf e}_x$, where
${\bf p}({\mathbf r}) = \sum_k \left[{\mathbf p}_k\, \delta({\mathbf
r}-{\mathbf r}_k) + \text{H.c.}\right]/2$ and $\sigma_z({\mathbf r})
= \sum_k \sigma_{z,\,k}\, \delta({\mathbf r}-{\mathbf r}_k)$ are the
momentum and spin density, respectively.

The hydrodynamic equations also  permit to calculate the relative
amplitudes of the density and spin density oscillations
characterizing the propagation of sound.  In terms of the magnetic
polarizability $\mathcal{M}$ we find
\begin{eqnarray}
\left(\frac{\delta s}{\delta n}\right)_{\text{II}} &\hspace{-1mm} =&
\frac{k_0 \mathcal{M}\cos \alpha}{1+ k_0^2 \mathcal{M}} \sqrt{
\frac{2 \left[ G_2 +G_1 \left(1+k_0^2 \mathcal{M} \right) \right]}{1
+k_0^2 \mathcal{M} \sin^2\alpha}} \qquad  \label{eq:amplitude_II}\\
&&+ \frac{\sqrt{1+\left(k_0^2-2 G_2 \right) \mathcal{M}}}{1+k_0^2
\mathcal{M}} \nonumber\\
\left(\frac{\delta s}{\delta n}\right)_{\text{III}} &\hspace{-1mm}
=& \frac{2k_0\mathcal{M} \cos\alpha \sqrt{G_1}}{\sqrt{2\left(1+k_0^2
\mathcal{M} \right) \left(1+k_0^2 \mathcal{M} \sin^2\alpha \right)}}
\label{eq:amplitude_III}
\end{eqnarray}
in phases II and III respectively. Equations (\ref{eq:amplitude_II})
and (\ref{eq:amplitude_III}) show that, near the transition between
phases II and III, the amplitude of the spin density fluctuations
$\delta s$ of the sound waves propagating along the $x$-direction
($\sin\alpha=0$) are strongly enhanced with respect to the density
fluctuations $\delta n$, as a consequence of the divergent behavior
of the magnetic susceptibility. This suggests that an effective way
to excite these phonon modes is through a coupling with the spin
degree of freedom as recently achieved in two-photon Bragg
experiments on Fermi gases \cite{Vale2012}. For sound waves
propagating in the direction orthogonal to $x$ the situation is
instead different. In particular in phase III sound waves are purely
density oscillations ($\delta s=0$).

A major usefulness of the hydrodynamic equations is that they can be
easily extended to trapped non-uniform configurations. In the
simplest $G_2 = 0$ case, corresponding to $G_1 = n g/2$, where the
wave vector ${\bf p}_1 = k_1 \hat{\mathbf e}_x$ is
density-independent, the chemical potential is given by the
Bogoliubov form $\mu=gn +\kappa $ with $\kappa$ independent of the
density, in both phase II and phase III, and the 3D hydrodynamic
equations can be reduced to the compact form:
\begin{equation}
\partial_t^2\delta n = g \left[\left(1-1/Z\right)\nabla_x \left(
n\nabla_x \delta n \right) +\nabla_\perp \left(n\nabla_\perp\delta
n\right)\right] \label{HDdeltan}
\end{equation}
Here $n$ is the Thomas-Fermi density profile given, in the presence
of harmonic trapping $V_{\text{ho}}({\bf r})= (\omega_x^2x^2+
\omega_y^2 y^2 + \omega_z^2 z^2)/2$, by an inverted parabola:
$n({\bf r}) = [\mu_0 - V_{\text{ho}}({\bf r})]/g$, with $\mu_0$
fixed by the normalization condition. One can easily check that all
the solutions holding for usual BECs \cite{stringari96} still hold
in the presence of spin-orbit coupling, with the simple replacement
of the trapping frequency $\omega_x$ with $\omega_x\sqrt{1-1/Z}$.
This reproduces the result
\begin{equation}
\omega^2_D=\frac{\omega^2_x}{1+k^2_0 \mathcal{M}}
\label{dipole}
\end{equation}
derived in \cite{trento2} for the frequency of the dipole
oscillation along the $x$-axis using a sum rule approach and also
shows that the frequency of the other hydrodynamic modes involving a
motion of the gas along the $x$ axis will be  quenched. The
quenching of the dipole mode due to spin-orbit coupling has been
recently observed in the experiment of \cite{Chen2012}.

\section{Conclusion}

In conclusion we have investigated the dynamic behavior of a
Bose-Einstein condensate with spin-orbit coupling, pointing out the
occurrence of novel features of high relevance for future
experiments, like the strong quenching exhibited by the sound
velocity near the second-order transition between the plane wave and
the zero momentum phases, the anisotropy of the compressibility, and
the occurrence of a roton minimum in the excitation spectrum. Our
theoretical predictions can be tested in future experiments bases on
two-photon Bragg spectroscopy and are expected to deeply influence
the superfluid behavior of the gas.

\acknowledgments

Useful discussions with G. Ferrari, G. Lamporesi, T. Ozawa, and I.
Spielman are acknowledged. This work has been supported by ERC
through the QGBE grant.

\appendix

\section{The coefficients in the response function}

The coefficients in the response function (\ref{eq:response}) can be
expressed as follows. In phase II we find
\begin{align*}
a &= \begin{aligned}[t]
&{} - \frac{q^4}{4} \\
&{} \hspace{-2mm} + \left[\left(k_0^2 + 3 k_1^2\right)\cos^2\alpha - 2 \left(k_0^2 - G_2\right) + 2 G_2 k_1^2/k_0^2\right]q^2 \\
&{} \hspace{-2mm} + 4\left(k_0^2 - 2 G_2\right) \left[\left(k_0^2 - k_1^2\right)\cos^2\alpha - k_0^2 + 2 G_2 k_1^2/k_0^2\right]
\end{aligned}\\
b_0 &= \begin{aligned}[t]
    &{} \frac{q^8}{16} - \left[\left(k_0^2 + k_1^2\right)\cos^2\alpha - k_0^2 - G_1 + G_2\right]\frac{q^6}{2} \\
    &{} \hspace{-2mm} + \big\{ \!\!\! \begin{aligned}[t]
            &{} \left(k_0^2 - k_1^2\right)^2\cos^4\alpha - 2\big[k_0^2\left(k_0^2 - k_1^2\right)\\
            &{} \hspace{5mm} + G_1\left(k_0^2 + 3 k_1^2\right) - G_2\left(k_0^2 - 5 k_1^2\right)\big]\cos^2\alpha\\
                    &{} + k_0^2\left(k_0^2 - 2 G_2\right) + 4 G_1\left(k_0^2 - G_2\right)\\
                    &{} + 2\left(k_0^2 - 2 G_1 - 2 G_2\right)G_2 k_1^2/k_0^2\big\}q^4
            \end{aligned}\\
    &{} \hspace{-2mm} - 8\left(k_0^2 - 2 G_2\right)\big[ \!\!\! \begin{aligned}[t]
                     &{} \left(k_0^2 - k_1^2\right)\left(G_1 + G_2 k_1^2/k_0^2\right) \cos^2\alpha\\
                     &{} \hspace{-1.4cm} - G_1 k_0^2- \left(k_0^2 - 2 G_1 - 2 G_2\right)G_2 k_1^2/k_0^2\big]q^2
            \end{aligned}
    \end{aligned} \\
b_1 &= \begin{aligned}[t]
    &{} q^4 + 4 \left[\left(k_0^2 - k_1^2\right)\cos^2\alpha + 2\left(G_1 + G_2\right)\right]q^2 \\
    &{} \hspace{-2mm} + 16 \left(k_0^2 - 2 G_2\right)\left(k_0^2 - k_1^2\right)G_2/k_0^2
       \end{aligned} \\
b_2 &= \begin{aligned}[t]
        &{} -\frac{q^4}{2} - 2\left[\left(k_0^2 - 3 k_1^2\right)\cos^2\alpha + k_0^2 + G_1 - G_2\right]q^2 \\
    &{} \hspace{-2mm} - 4\left(k_0^2 - 2 G_2\right)\left(k_0^2 - 2 G_2 k_1^2/k_0^2\right)
       \end{aligned}
\end{align*}
In phase III we instead obtain the results
\begin{align*}
a &= \begin{aligned}[t]
        &{} - \frac{q^4}{4} - \left(\Omega - k_0^2\cos^2\alpha + 2 G_2\right)q^2 \\
    &{} \hspace{-2mm} - \Omega\left[\Omega - 2 \left(k_0^2\cos^2\alpha - 2 G_2\right)\right]
       \end{aligned}\\
b_0 &= \begin{aligned}[t]
    &{} \frac{q^8}{16} + \left[\Omega - 2\left(k_0^2\cos^2\alpha - G_1 - G_2\right)\right]\frac{q^6}{4}\\
    &{} \hspace{-2mm} + \big[\begin{aligned}[t]
            &{} \Omega^2 - 4\left(k_0^2\cos^2\alpha - 2 G_1 - G_2\right)\Omega \\
            &{} + 4\left(k_0^2\cos^2\alpha - 2 G_1\right)\left(k_0^2\cos^2\alpha - 2 G_2\right)\big]\frac{q^4}{4}\\
            \end{aligned}\\
    &{} \hspace{-2mm} + 2 G_1 \Omega \left[\Omega - 2 \left(k_0^2\cos^2\alpha - 2 G_2\right)\right]q^2
    \end{aligned}\\
b_1 &= 0\\
b_2 &= \begin{aligned}[t]
        &{} -\frac{q^4}{2} - \left[\Omega + 2\left(k_0^2\cos^2\alpha + G_1 + G_2\right)\right]q^2 \\
    &{} \hspace{-2mm} - \Omega\left(\Omega + 4 G_2\right)
       \end{aligned}
\end{align*}


\begin{thebibliography}{99}
\bibitem{note_g}
For the spin asymmetric case where $g_{\uparrow\uparrow}\neq
g_{\downarrow\downarrow}$, one can define the interaction parameters
as $G_1=n(g_{\uparrow\uparrow}+g_{\downarrow\downarrow} + 2
g_{\uparrow\downarrow})/8$, $G_2=n(g_{\uparrow\uparrow}
+g_{\downarrow\downarrow} - 2 g_{\uparrow\downarrow})/8$, and
$G_3=n(g_{\uparrow\uparrow} -g_{\downarrow\downarrow})/4$. By tuning
the parameter $\delta$, the asymmetric effect caused by $G_3$ can be
compensated. The ground state then remains the same as in the spin
symmetric case \cite{Li2012}.
\bibitem{sigmaz}
Notice that result (\ref{eq:sigma_x}) for $\langle\sigma_x\rangle$
holds in the spin rotated frame where the Hamiltonian takes the form
(\ref{eq:Ham}). Since the operators $\sigma_x$ and $\sigma_z$ do not
commute, the average value of $\sigma_x$ evaluated in spin space,
calculated in the original frame, exhibits an additional oscillatory
behavior $\langle\sigma_x\rangle\cos(2k_0x-\Delta\omega_Lt) -
\langle\sigma_y\rangle\sin(2k_0x-\Delta\omega_Lt)$, with
$\langle\sigma_x\rangle$ and $\langle\sigma_y\rangle$ given by
(\ref{eq:sigma_x}) and (\ref{eq:sigma_y}), characterizing the laser
potential of Eq.~(\ref{eq:h00}) (an analogous result holds for the
average value of $\sigma_y$).
\bibitem{note3}
The equation for the spin density takes the form $\partial s
/\partial t = k_0 \nabla_x n - \nabla\cdot [s \nabla
(\varphi_\uparrow + \varphi_\downarrow ) + n \nabla
(\varphi_\uparrow - \varphi_\downarrow )]/2 - \Omega \sqrt{n^2 -
s^2} \sin(\varphi_\uparrow - \varphi_\downarrow)$ showing that the
difference ($\varphi_\uparrow-\varphi_\downarrow$) between the two
phases should be small if the frequency of the oscillation is much
smaller than the Raman coupling  $\Omega$. The oscillation of the
relative phase plays a crucial role in the excitation of the upper
branch, whose frequency is of the order of $\Omega$.
\end{thebibliography}
\end{document}